\documentclass[12pt,preprint]{aastex}

\def\zabs{$z_{\rm abs}$}

\def\h2{H$_2$~}

\def\kms{km~s$^{-1}$}

\makeatother

\begin{document}

%%%\twocolumn[%
%%%
%%%\submitted{Submitted to ApJL}

\title{Detecting cold gas at intermediate redshifts: GMRT survey using Mg II systems.}

\author{Neeraj Gupta$^1$, R. Srianand$^2$, P. Petitjean$^3$, P. Khare$^4$, D. J. Saikia$^1$,  D. G. York$^5$}

\begin{abstract}
Intervening H~{\sc i} 21-cm absorption systems at $z \ge 1.0$ are very rare and only 
4 confirmed detections have been reported in the literature. Despite their scarcity, 
they provide interesting and unique insights into the physical conditions in the 
interstellar medium of high-$z$ galaxies.  Moreover, they can provide 
independent constraints on the variation of fundamental constants. We report 3 new 
detections based on our ongoing Giant Metrewave Radio Telescope (GMRT) survey 
for 21-cm absorbers at $1.10\le z_{\rm abs}\le 1.45$ from candidate damped
Lyman-$\alpha$ systems. The 21-cm lines are narrow for the  \zabs = 1.3710 system towards 
SDSS J0108$-$0037  and  \zabs = 1.1726 system toward SDSS J2358$-$1020.
Based on line full-width at half maximum, the kinetic temperatures are $\le 5200$ K 
and $\le 800$ K, respectively. The 21-cm absorption profile of the third system,  
\zabs =1.1908 system towards SDSS J0804+3012, is shallow, broad and complex, extending 
up to 100 \kms.  The centroids of the 21-cm lines are found to be shifted with 
respect to the corresponding centroids of the metal lines derived from SDSS spectra.  
This may mean that the 21-cm absorption is not associated with  the strongest metal line component.
\end{abstract}

%%%]
\altaffiltext{1}{NCRA, Postbag 3, Ganeshkhind, Pune 411007, India; 
neeraj@ncra.tifr.res.in; djs@ncra.tifr.res.in}
\altaffiltext{2}{IUCAA, Postbag 4, Ganeshkhind, Pune 411007, India; anand@iucaa.ernet.in}
\altaffiltext{3}{Institut d'Astrophysique de Paris -- CNRS, 98bis Boulevard
Arago, F-75014 Paris, France; ppetitje@iap.fr}
\altaffiltext{4}{Department of Physics, Utkal University, Bhubaneswar 751004,
India; khare@iopb.res.in}
\altaffiltext{5}{Department of Astronomy \& Astrophysics and Enrico Fermi
Institute, 5640 S. Ellis Avenue, Univ. of Chicago, Chicago, IL 60637; don@oddjob.uchicago.edu}

\section{Introduction}

The diffuse interstellar medium exhibits a wide range of physical
conditions such as temperature, density and radiation field that are
influenced by in-situ star-formation, cosmic ray energy density,
photoelectric heating by dust as well as mechanical energy input from
both impulsive disturbances such as supernova explosions and
steady injection of energy in the form of stellar winds.  Therefore,
understanding the physical conditions in the gas and the processes that
maintain these is important for understanding galaxies and their
evolution.  The damped Lyman-$\alpha$ systems (DLAs), 
with log~$N$(H~{\sc i})$\ge$20.3, are a major
reservoir  of H~{\sc i} at high $z$ and possibly the progenitors of
present-day galaxies (see Wolfe et al. 2005). At
high $z$, despite many attempts, only a handful of DLA galaxies have been
detected based on line and/or continuum emission  (see M\o ller et al. 2004).

\begin{figure}
\centering
\vbox{
\centering
\includegraphics[width=9.cm,height=6cm,angle=0]{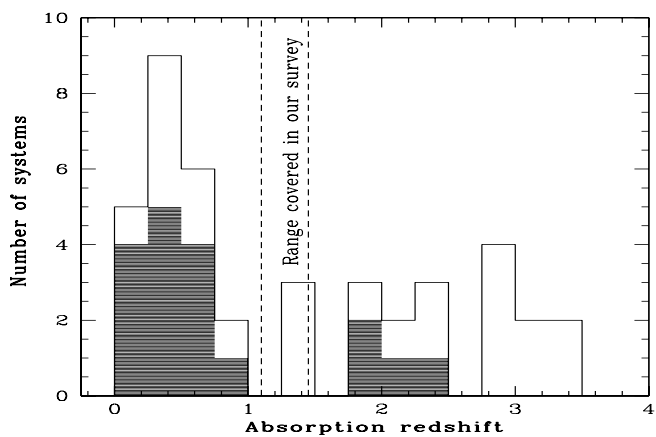}}
\caption{Line-histogram is the redshift distribution of the DLAs for
which redshifted 21-cm observations are reported (Table.~1 of Curran et al. 2005
and $z= 2.347$ towards PKS 0438$-$436 from Kanekar et al. 2006).
Shaded histogram is for the confirmed 21-cm absorption systems. 
The redshift range covered in our GMRT survey is also marked. 
}
\label{Fig1}
\end{figure}

Our understanding of  physical conditions in DLAs at $z\ge1.8$  is
based primarily on the absorption lines of  H$_2$ molecules and atomic 
fine-structure lines. A systematic search for H$_2$ in DLAs at ($z_{\rm abs}>1.8$) 
has resulted in a detection in $\sim$15\% of the cases (see Ledoux et al.
2003).  Usually, the DLAs with H$_2$ absorption also show absorption
lines of C~{\sc i}, C~{\sc i}$^*$, C~{\sc i}$^{**}$ and C~{\sc ii}$^*$.
Detailed  investigations show that H$_2$ components have properties similar to 
that of the cold neutral medium (CNM) in a radiation field of moderate 
intensity originating from local star formation activity (Srianand et al. 2005).
Using C~{\sc ii$^*$} absorption lines Wolfe et al. (2004) have concluded that 
most of the DLAs at high $z$ consist of CNM gas.  However, above techniques 
cannot be used to probe nature of the gas at low and intermediate redshifts.
In this case detecting 21-cm absorption line is one complementary
way to probe the nature of absorbing gas.  There
have been several systematic searches for 21-cm absorption in DLAs
undertaken by various groups (Briggs \& Wolfe 1983;  Lane 2000; Kanekar
\& Chengalur 2003, Curran et al. 2005) with limited success.  
To date, in the literature 38 DLAs have been searched for 21-cm absorption,
resulting in 17 detections most of which occur $z<1$ (see Fig.~\ref{Fig1}). 
%The 
%detection seems to be more frequent at the low $z$.  
The column density of 
H~{\sc i} for an optically thin cloud that covers a fraction $f_c$ of the 
background source is related to the optical depth $\tau(v)$
in a velocity interval $v$ and $v+$d$v$ and to the spin temperature ($T_s$) by
\begin{equation}
N{\rm(H~I)}=1.835\times10^{18}~{T_s\over f_c}\int~\tau(v)~{\rm d}v~{\rm cm^{-2}}.
\label{eq1}
\end{equation}
The low detection rate at high $z$ can therefore be attributed
either to the gas being warm (high $T_s$) or to a low  value of the covering 
factor ($f_c$) through high $z$ geometry effects (see Kanekar \& Chengalur, 
2003 and Curran \& Webb, 2006, respectively). However, the redshift coverage 
is sparse and measurements are available only for a few systems at $1\le z\le 2$ 
(Fig.~\ref{Fig1}). To improve the statistics we have started
a systematic survey of 21-cm absorption at 1.10$\le z\le$1.45
using the 610-MHz receiver at GMRT. Here, we report the results from
the first phase of this survey that has resulted in 3 new detections.

Obtaining a new DLA sample at $z\le 1.6$ is virtually impossible in the
absence of UV-spectrographs in space as the redshifted wavelength of
Lyman-$\alpha$ line falls below the atmospheric cutoff.  However,  
Rao \& Turnshek (2000) have shown that the DLAs can be preselected on 
the basis of equivalent widths of Mg~{\sc ii}, Fe~{\sc ii} and Mg~{\sc i} 
absorption lines. Specifically, they found that 50\% of the absorbers with 
rest frame W(Fe~{\sc ii}$\lambda$2600) and W(Mg~{\sc ii}$\lambda$2796) 
greater than 0.5 \AA~were confirmed DLAs. The detection rate becomes 100\%  
when W(Mg~{\sc i})$\ge$ 0.5 \AA. It is also clear from Fig.~3.4 of 
Lane (2000) that such a selection will also ensure that 50\% of these 
candidates are detected in 21-cm absorption.
Motivated by this, we have begun a GMRT 21-cm survey of DLA  candidates selected on
the basis of  W(Mg~{\sc ii}) in the redshift range 1.10$\le z \le 1.45$.
Our complete sample is drawn from the catalog of 7421 Mg~{\sc ii} systems
with W(Mg~{\sc ii})$\ge1$\AA~and 0.3$\le z\le$1.9,  detected along the
line of sight toward 45,023 QSOs in SDSS DR3 (Prochter, Prochaska \&
Burles, 2006).  There are 2857 systems at 1.10$\le z \le 1.45$.  
Out of these, we have selected all the 26 systems that have 
an estimated 610 MHz flux density in excess of 100 mJy from 
NVSS and First catalog. These form our main sample. 
In addition to this we have also observed three other
sources [J0108$-$0037 (York et al. 2006), 
J0240$-$2309 and J1604$-$0019 (Lanzetta, Wolfe \& Turnshek, 1987)] 
that have total
flux density at 610 MHz in excess of 1 Jy and have strong Mg~{\sc ii} systems.

\section{Observations and Results}

\begin{figure}
\centering
\vbox{
\centering
\includegraphics[bb=17 180 592 462,width=9.cm,height=3.cm,angle=0,clip=true]{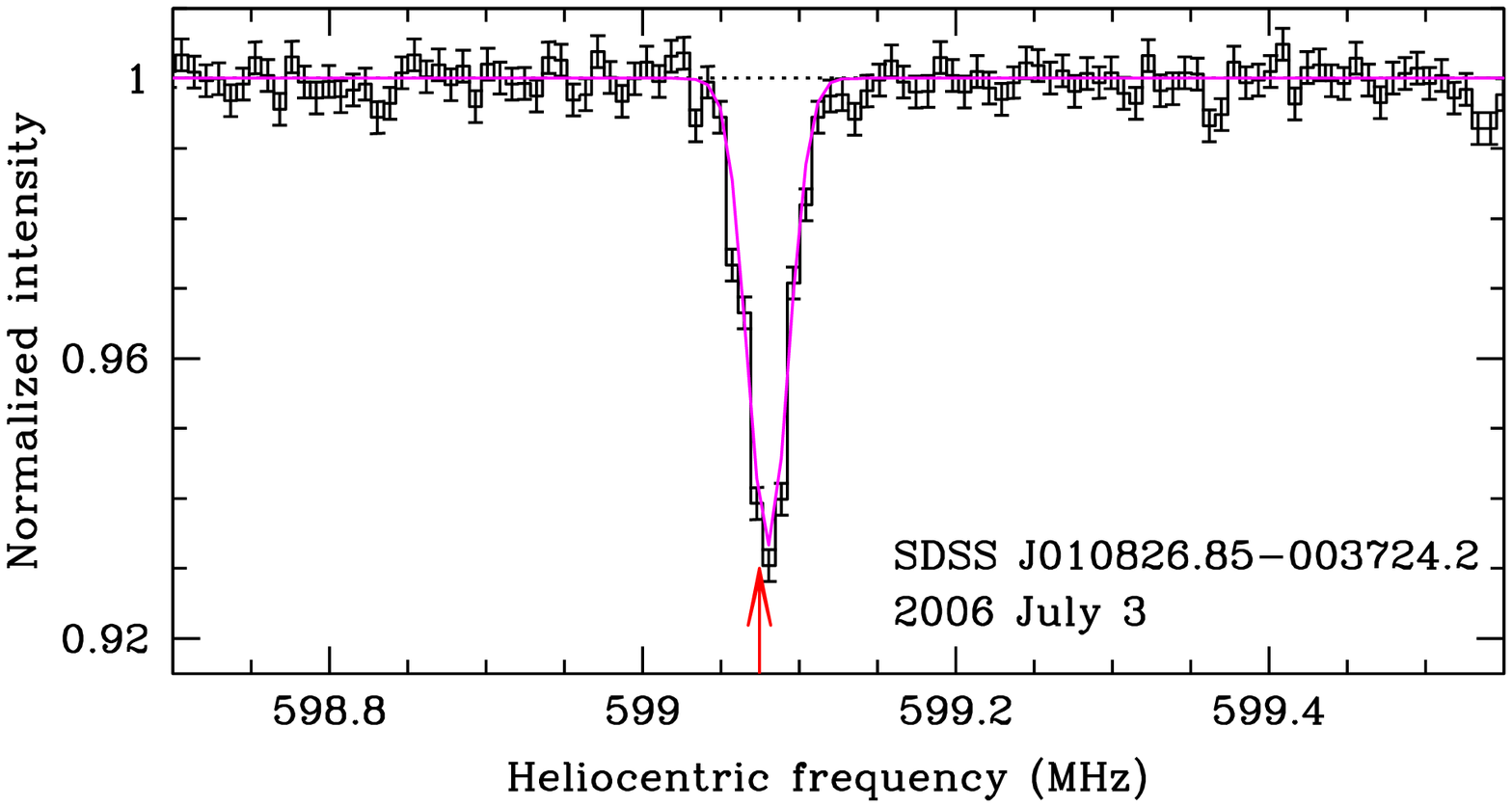}
\includegraphics[bb=17 176 592 718,width=9.cm,height=5cm,angle=0,clip=true]{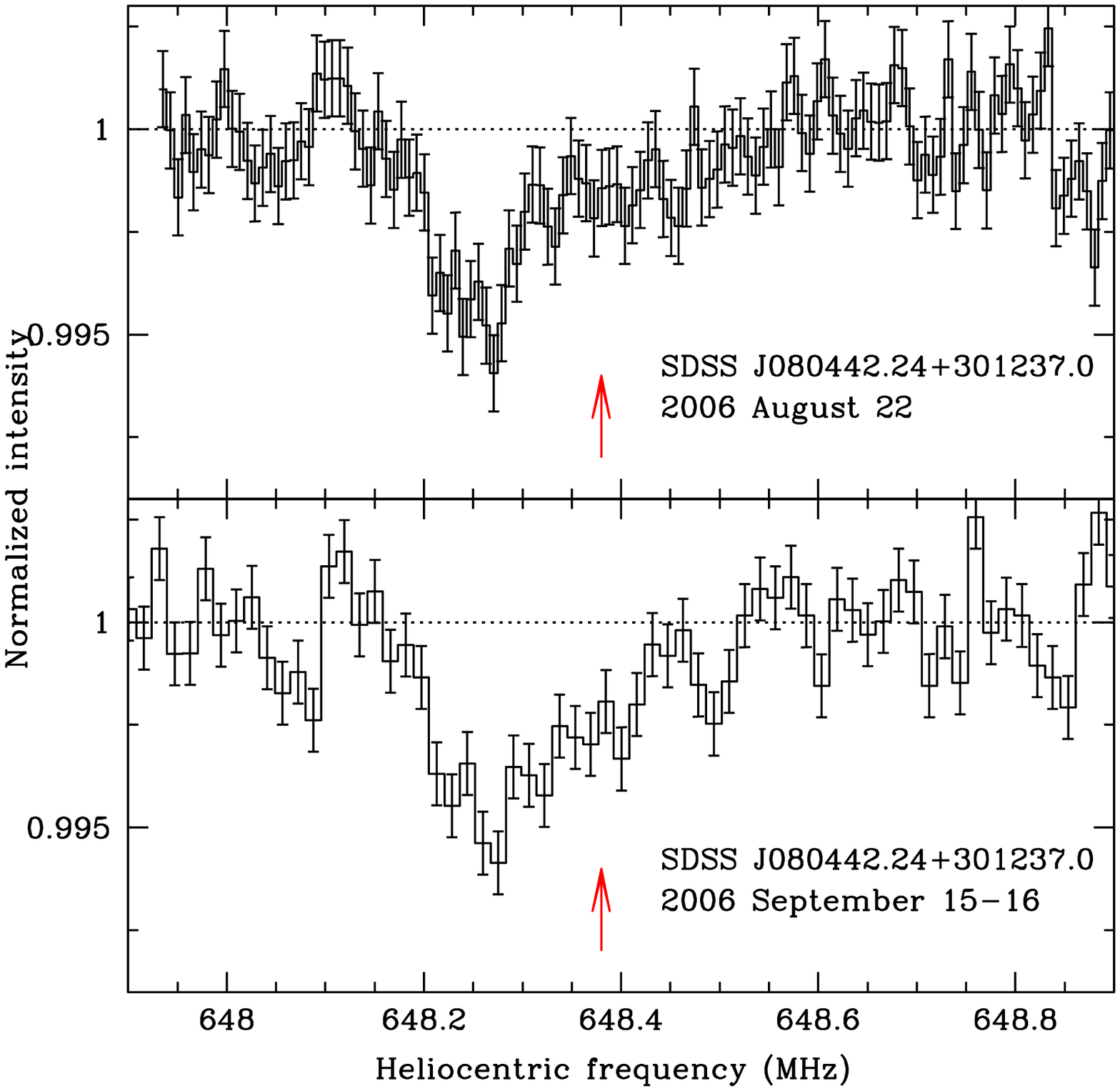}
\includegraphics[width=9.cm,height=5cm,angle=0]{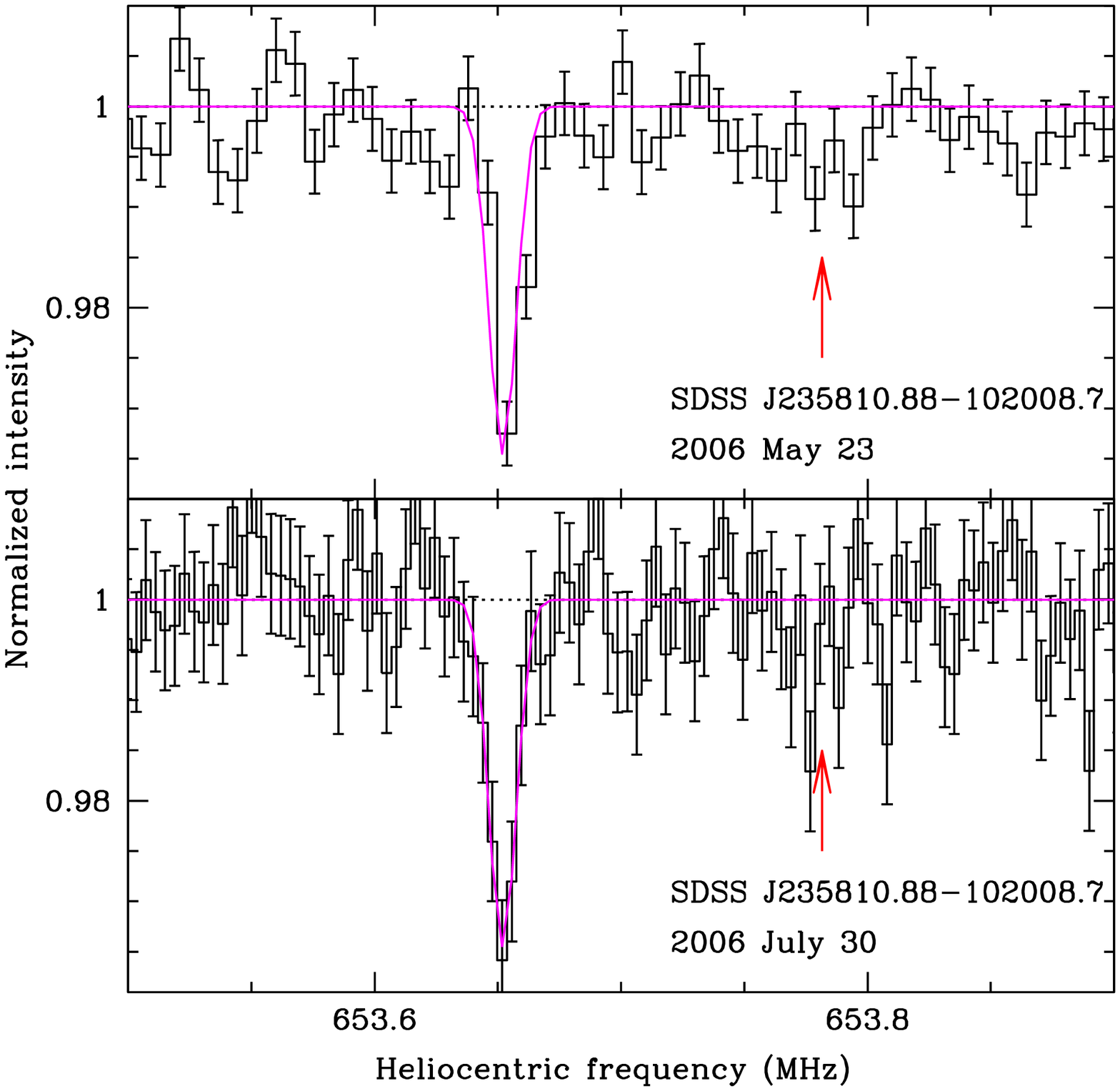}
}
\caption{GMRT H~{\sc i} spectra of the sources with 21-cm absorption.  
Single Gaussian fits are overplotted in the case of
J0108$-$0037 and J2358$-$1020.  Arrows mark the expected positions of
21-cm absorption based on the metal absorption lines.}
\label{fig_det}
\end{figure}

\begin{figure}
\centering
\vbox{
\centering
\includegraphics[bb=17 180 572 458,width=9.cm,height=2.25cm,angle=0,clip=true]{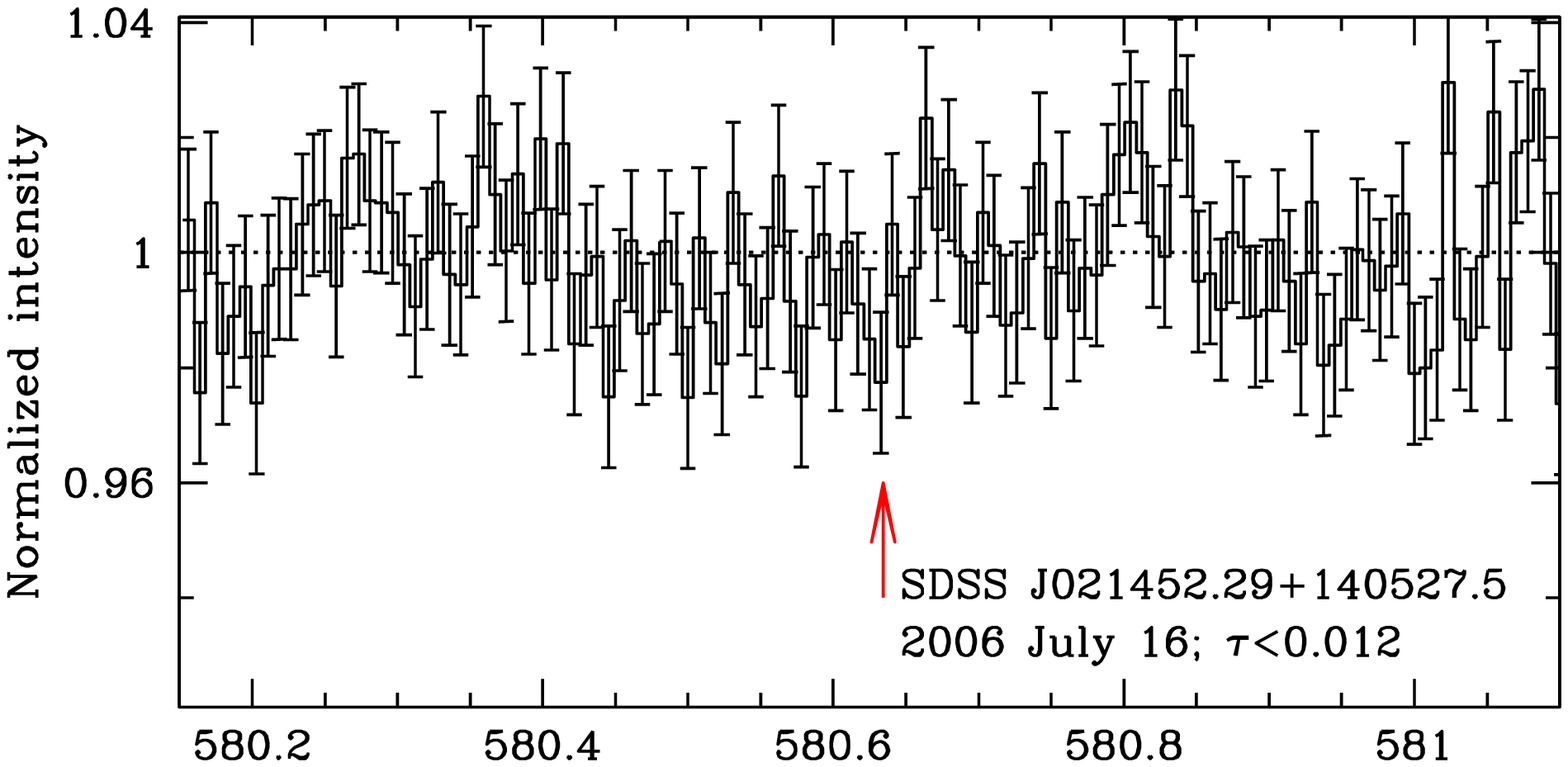}
\includegraphics[bb=17 180 572 458,width=9.cm,height=2.25cm,angle=0,clip=true]{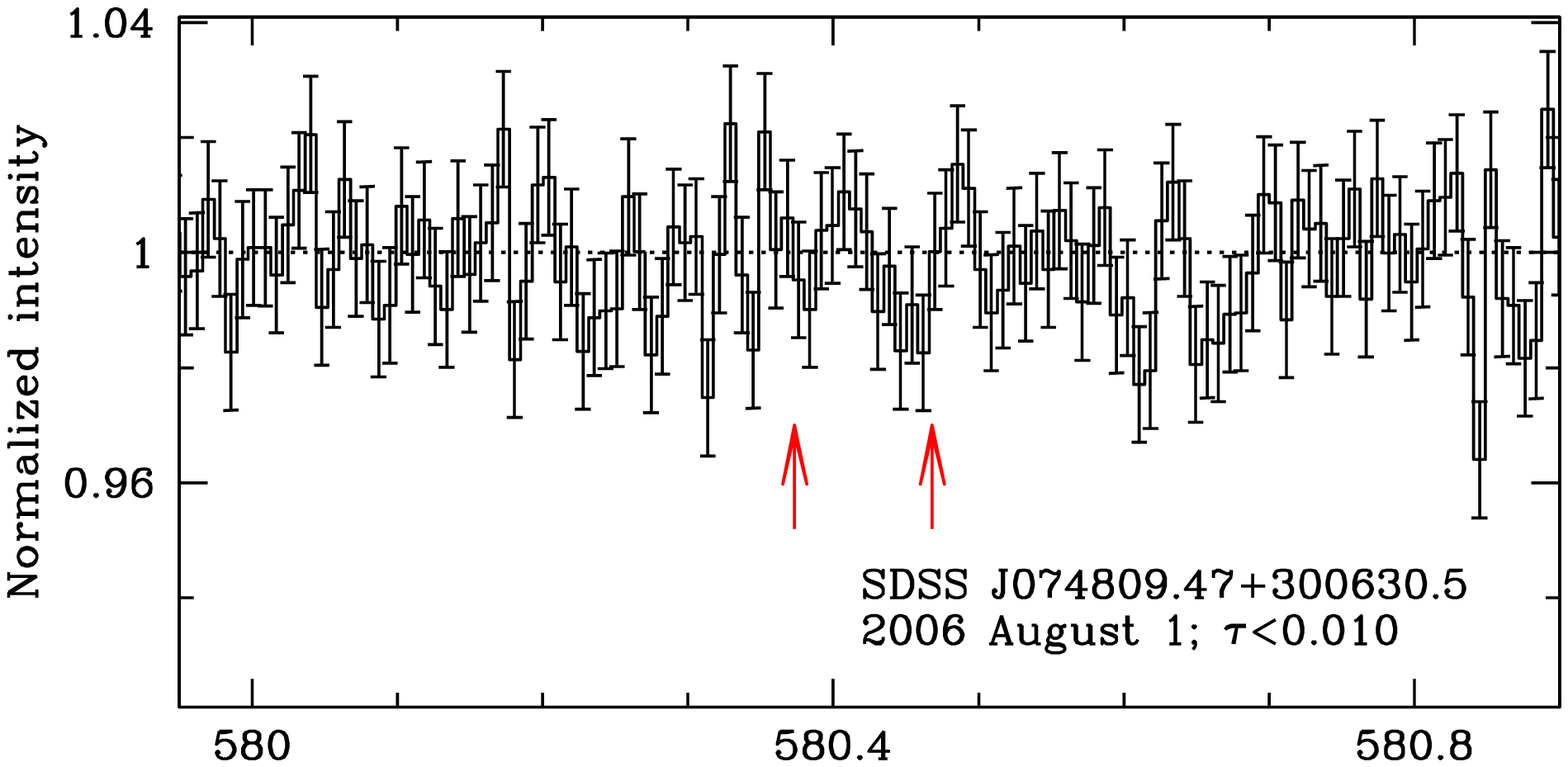}
\includegraphics[bb=17 180 572 458,width=9.cm,height=2.25cm,angle=0,clip=true]{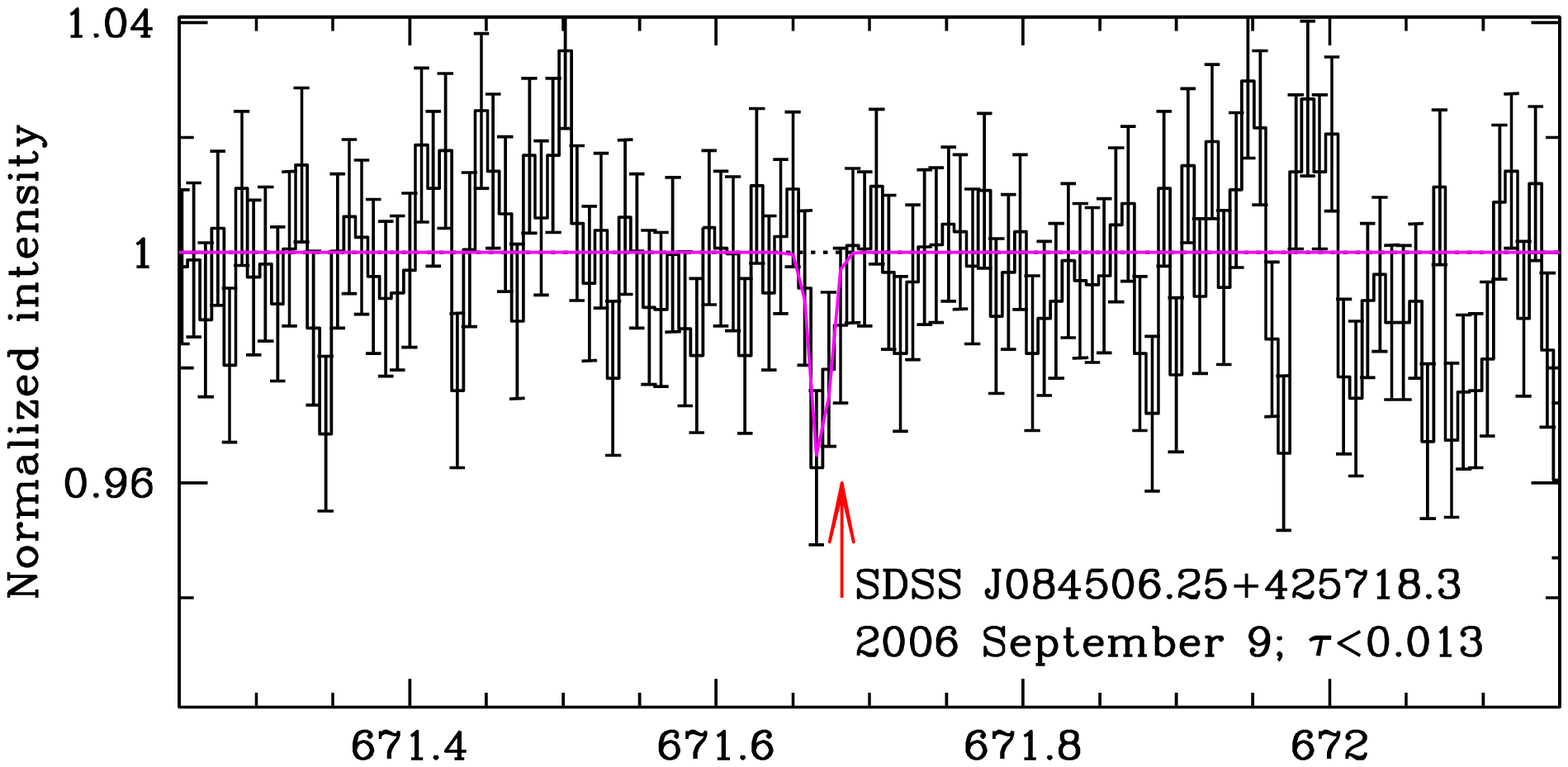}
\includegraphics[bb=17 180 572 458,width=9.cm,height=2.25cm,angle=0,clip=true]{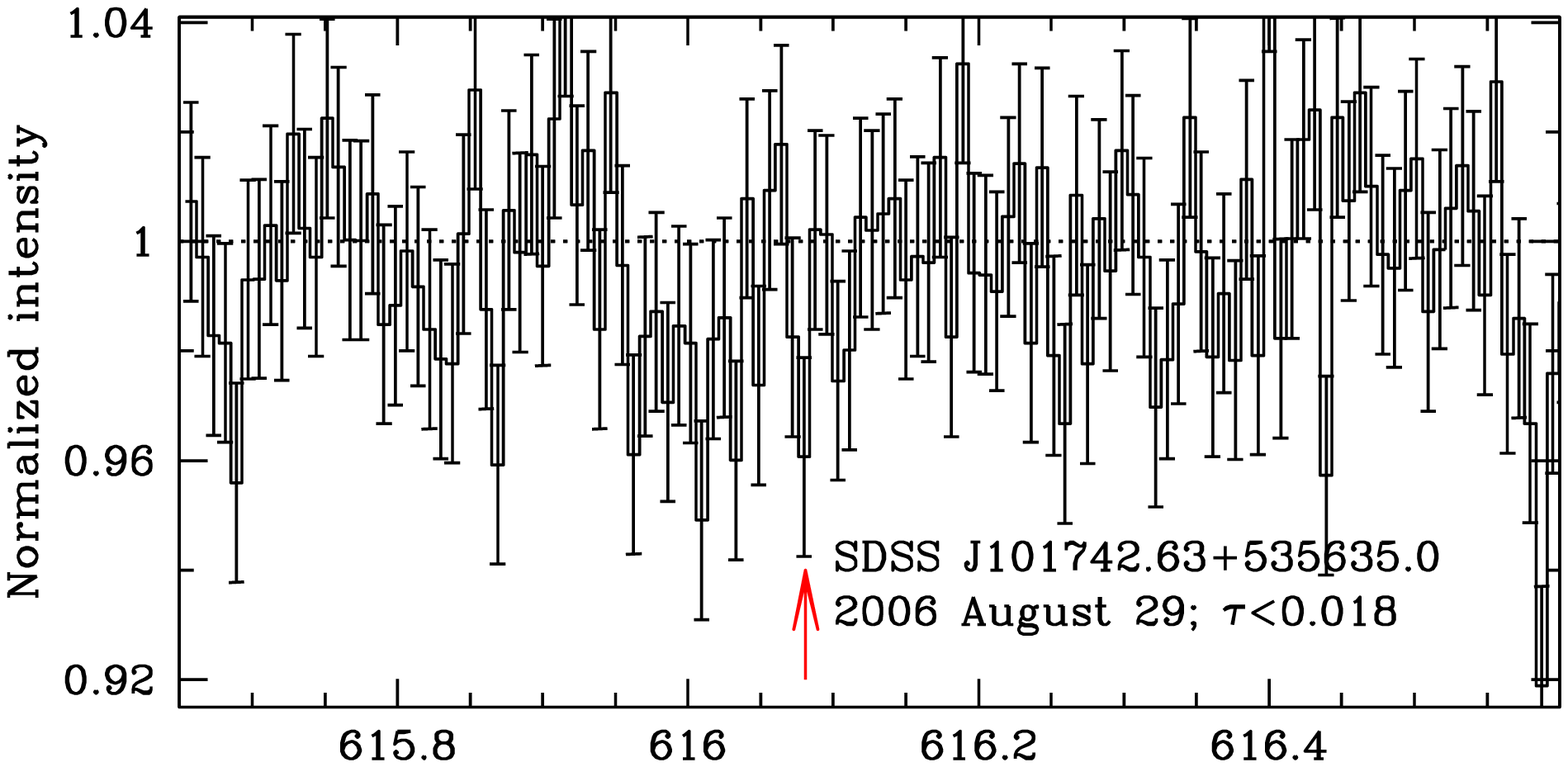}
\includegraphics[width=9.3cm,height=4cm,angle=0]{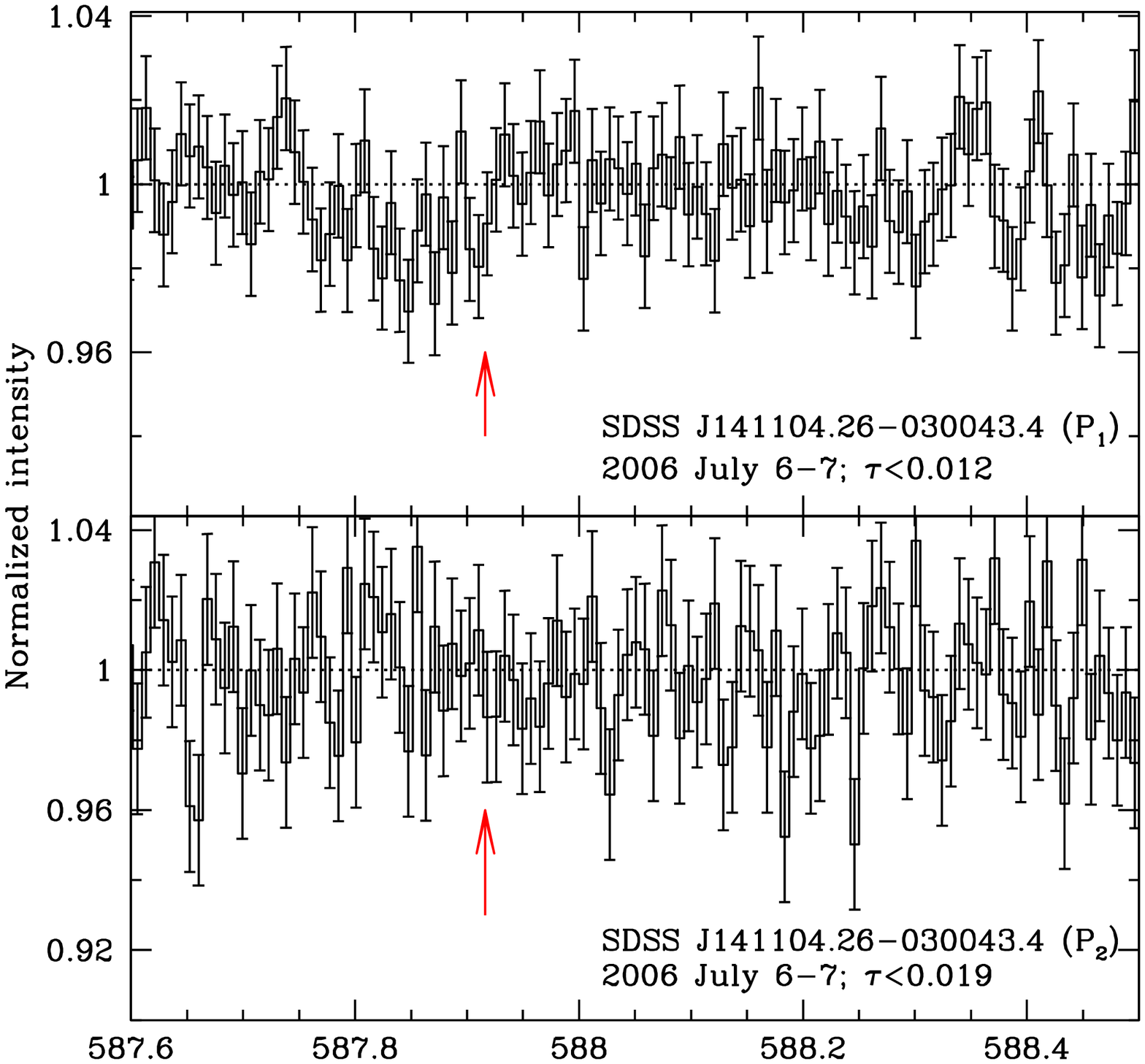}
\includegraphics[bb=17 146 572 458,width=9.cm,height=2.5cm,angle=0,clip=true]{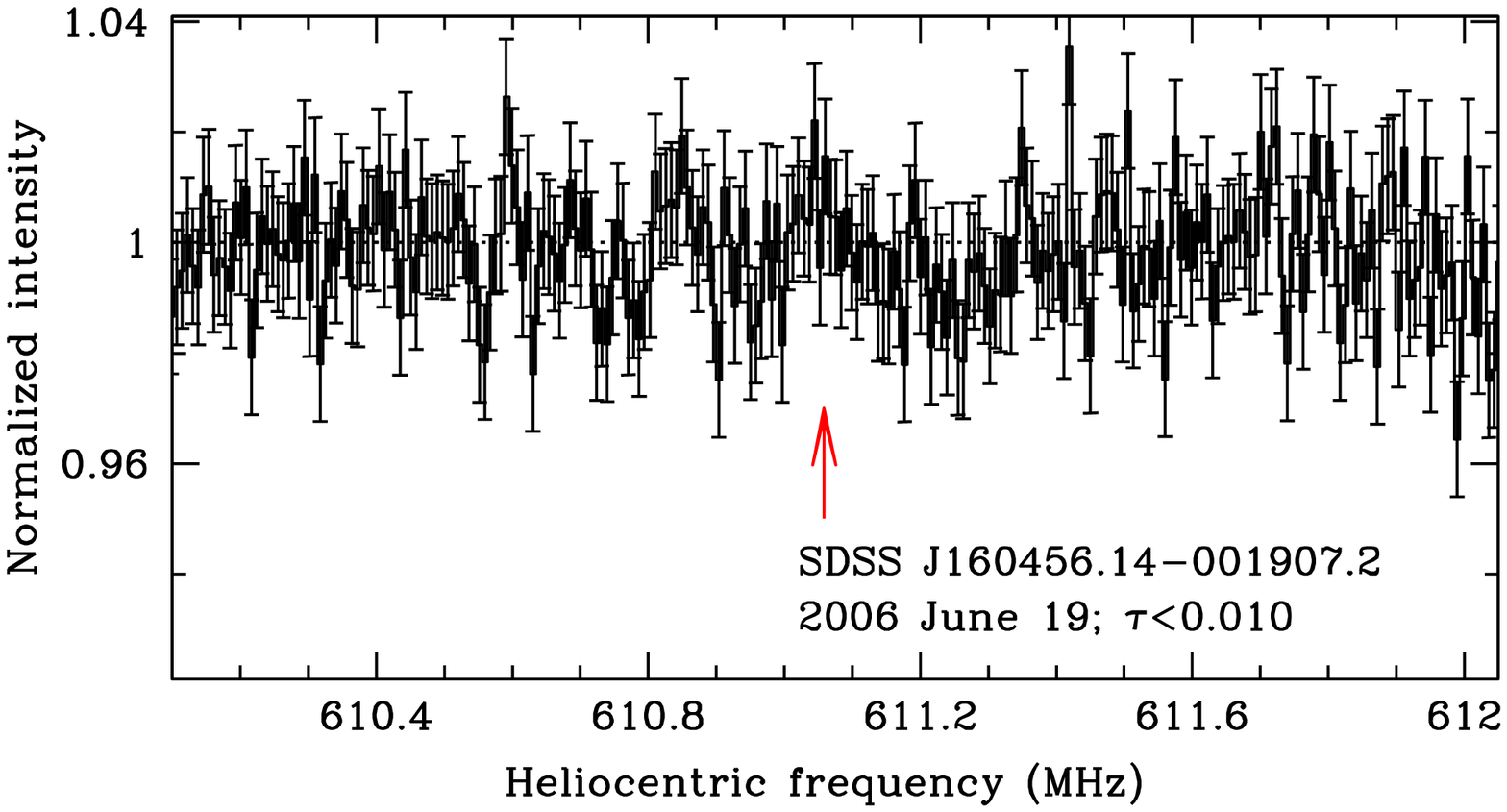}
}
\caption{Spectra for the sources with non-detection in 21-cm
absorption.  Arrows mark the expected position of
21-cm absorption based on the metal absorption lines.}
\label{fig_nondet}
\end{figure}

\begin{figure}
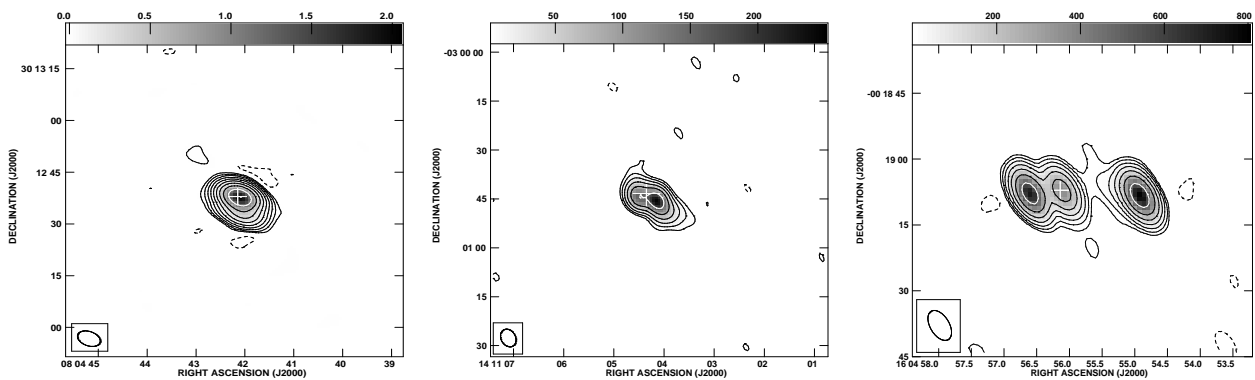

\centering
\hbox{
\centering
\includegraphics[width=5.0cm,height=5.5cm,angle=270]{f4a.ps}
\includegraphics[width=5.0cm,height=5.5cm,angle=270]{f4b.ps}
\includegraphics[width=5.0cm,height=5.5cm,angle=270]{f4c.ps}
}
\caption{GMRT maps of sources resolved in our observations with contour levels as 
n$\times$($-1$,1,2,4,...) mJy/beam. 
{\bf (a)} J0804+3012 with an rms of 0.84 mJy/beam and beam of
6.36$^{\prime\prime}\times$4.09$^{\prime\prime}$ with position angle (PA)= 
67$^\circ$ and n=4. 
{\bf (b)} J1411$-$0300 with an rms of 1.25 mJy/beam  and beam
of 5.67$^{\prime\prime}\times$4.32$^{\prime\prime}$ with PA=28$^\circ$ and n=5.  
{\bf (c)} SDSS J1604$-$0019 with an rms of 1.50 mJy/beam  and
beam of 7.55$^{\prime\prime}\times$4.34$^{\prime\prime}$ with PA=32$^\circ$ 
and n=8. The position of the optical source in each image is marked as a cross.}
\label{fig_map}
\end{figure}

In the first phase of our GMRT survey we have observed 10 systems 
 (Table~\ref{obslog}) from our sample.  Usually, a 1 MHz
bandwidth  split into 128 frequency channels was used to acquire data in
the two circular polarization channels RR and LL. Broader (2 MHz for
J0804+3012) or narrower (0.5 MHz for  J2358$-$1020) bandwidths were used in
the subsequent observing runs that were carried out for confirming the
detected absorption.  The  data were reduced  using the Astronomical
Image Processing System (NRAO AIPS package) following standard procedures
as described in Gupta et al. (2006). Final spectra for the three new
detections and six non-detections  are  presented in  Figs.~\ref{fig_det}
and  \ref{fig_nondet} respectively.  Analysis of \zabs = 1.3647
towards J0240$-$2309 is presented in Srianand, Gupta \& Petitjean
(2006). 
In most of the cases the background quasar is unresolved for the typical
5$^{\prime\prime}\times$5$^{\prime\prime}$ synthesized beam achieved in
our observations.  However, for J0804$+$3012, J1411$-$0300 and J1604$-$0019  
the radio sources are extended (see Fig.~\ref{fig_map}).
For J0804+3012, 21-cm absorption is detected in the spectrum extracted 
towards the radio peak (Fig.~\ref{fig_det}).  Higher spatial resolution as 
well as S/N observations will be required to investigate its variation across the source.  
In the case of J1411$-$0300, the radio peak is not consistent with the location of 
the optical source. 21-cm absorption is not detected towards either the 
strongest radio peak (P1) or another  peak (P2) northeast to P1 
(Figs.~\ref{fig_nondet} and ~\ref{fig_map}).  No 21-cm absorption
is detected  towards any of the three prominent radio peaks seen in 
J1604$-$0019. Fig.~\ref{fig_nondet} shows the spectrum towards the central 
component. A narrow absorption  is present with $\tau_{\rm peak}\sim$0.04$\pm$0.01 
and FWHM=6.1$\pm$2.3 \kms~near the expected frequency towards J0845$+$4257 
(see Fig.~\ref{fig_nondet}).  The feature is present  in both the polarizations 
and in the spectra extracted using different baselines and time ranges.  At this 
stage we consider this as a tentative detection. 
 
\subsection{\zabs =1.3710 system towards J0108$-$0037}
The 21-cm absorption profile is well fitted  with a single Gaussian  having 
$\tau_{\rm peak}$   = 0.068$\pm$0.002 and FWHM = 15.4$\pm$0.6 \kms. 
From the observed optical depth we get the integrated column density,
$N$(H~{\sc i}) = 2.1$\times$10$^{18}$($T_{s}/f_c$) cm$^{-2}$.
High frequency VLBA observations suggest structures extending over 
60 mas ($\sim 50$ pc; H$_o$=71 km s$^{-1}$ Mpc$^{-1}$, $\Omega_m$=0.27,
$\Omega_\Lambda$=0.73)  with significant flux density in extended
emission (Beasley et al. 2002). Thus $f_c$ is uncertain. 
Assuming thermal broadening, gives a gas kinetic temperature
$T_K={\rm21.86\times ({FWHM\over km~s^{-1}})^2}\le$ 5200 K.  This is an 
upper limit as various other effects also could contribute to the line 
width. Now we can assume $T_s=T_K$ as is the case for thermodynamic 
equilibrium. If the gas is a typical CNM  (say $T_s=T_K = 100$ K) then 
log~[$f_c$$N$(H~{\sc i})] = 20.30. Thus, $N$(H{~\sc i}) in 
the 21-cm component is consistent with it being a DLA.
The 21-cm centroid is within 4 \kms~to that expected from the 
metal lines detected in the SDSS spectrum. The Zn~{\sc ii}+Cr~{\sc ii} 
blend at $\lambda$ = 2062 \AA, the Si~{\sc ii}$\lambda$1808 and the 
Mn~{\sc ii} lines are clearly detected even in the low-dispersion 
SDSS spectrum.  

%%%%%%
\subsection{\zabs =1.1908 system towards J0804+3012}
21-cm absorption is clearly detected towards J0804+3012 which is partially 
resolved in our observations (Fig.~\ref{fig_map}).
The absorption profile is consistent with a main component (with FWHM $\sim$ 
$44\pm19$ \kms and $\tau_{\rm peak}=0.006\pm0.001$) and a broad wing (extending 
up to $\sim 100$ \kms) towards higher frequency.  Our two-epoch observations 
with different spectral resolution confirm the main component which also 
shows the expected Doppler shift. However, deeper observations are required 
to ascertain the strength of the wing.It is interesting to note that the expected 
position of 21-cm absorption based on the SDSS spectrum coincides with the 
wing of the 21-cm absorption and not the stronger main component.  Based on the
profile obtained on 15$-$16 September 2006, we get an integrated column
density of $N$(H~{\sc i})=6.5$\times$10$^{17}$ ($T_{\rm s}$/$f_c$)
cm$^{-2}$.  Unlike the other two 21-cm absorbers, we do not detect Si~{\sc ii}$\lambda$1808 
(with W$\le$0.3\AA) and Zn~{\sc ii}$\lambda$2026 lines (with W$\le$0.2\AA) in the 
SDSS spectrum.  
   
%%%%%%
\subsection{\zabs =1.1726 system towards J2358$-$1020} 

Relatively weak 21-cm absorption is clearly detected and the profile can be well 
fitted with a single Gaussian component with $\tau_{\rm peak}$= 0.035$\pm$0.004 and
FWHM = 6.0$\pm$0.8 \kms(Fig~\ref{fig_det}). This implies $T_{\rm k}\le$800 K and log
[$f_c$ $N$(H~{\sc i})]$\le$ 20.5 in the 21-cm absorbing component. 
If the absorbing gas happens to be a CNM then log [$f_c$ $N$(H~{\sc i})] = 19.6.  
The value of $f_c$ could be close to unity as VLBA observations at 5 GHz by 
Fomalont et al. (2000) show the source to be $\le0.5$ mas, which corresponds to 
a projected size of $\le 4$ pc at \zabs =1.1726. Thus the inferred range in 
$N$(H~{\sc i}) is consistent with the 21-cm absorbing gas being a sub-DLA.  
Like the previous system, the 21-cm absorption is shifted by 70 \kms~with 
respect to the optical redshift based on the SDSS data. The 
Zn~{\sc ii}+Cr~{\sc ii} blend at $\lambda$ = 2062 \AA~is clearly detected in 
the SDSS spectrum.  The compactness of the background radio source and the 
simplicity of the 21-cm absorption profile make this system 
an ideal case for probing the variation of fundamental constants.

\section{Discussion}

In this letter we have reported the detection of 21-cm absorption in 
three systems. This has substantially increased the number of 21-cm 
absorption systems at $z>1.0$, as prior to this only four systems 
were known.
In two of our systems (\zabs = 1.1908 towards J0804$+$3012 and 
\zabs = 1.1726 towards J2358$-$1020) the centroid  of
the strongest 21-cm absorption component is  shifted with 
respect to the centroid of metal lines derived from SDSS spectra.
This may mean that the strongest 21-cm absorption component is not 
associated with the strongest metal absorption component.  Such 
shifts have already been observed in other systems (see 
Srianand et al. 2005; Kanekar et al. 2006; Heinm\"uller et al. 2006).  
It is also known that \h2 carrying components in DLAs, that are 
believed to trace the CNM gas, often show similar shifts 
(although less pronounced) with respect to the strong metal line components 
(Ledoux et al. 2003. H$_2$ components often distinguish themselves 
with respect to the rest of the velocity components by the presence 
of excited fine-structure lines of C~{\sc i} and C~{\sc ii} and 
in some cases by large depletion factors  (e.g. Rodr\' iguez et al. 2006). 
It would be very important to obtain high resolution optical spectra of 
our sources to see if this is also true in the case of the 21-cm absorption
systems. 

\begin{figure}
\centering
\vbox{
\centering
\includegraphics[width=9.cm,angle=0]{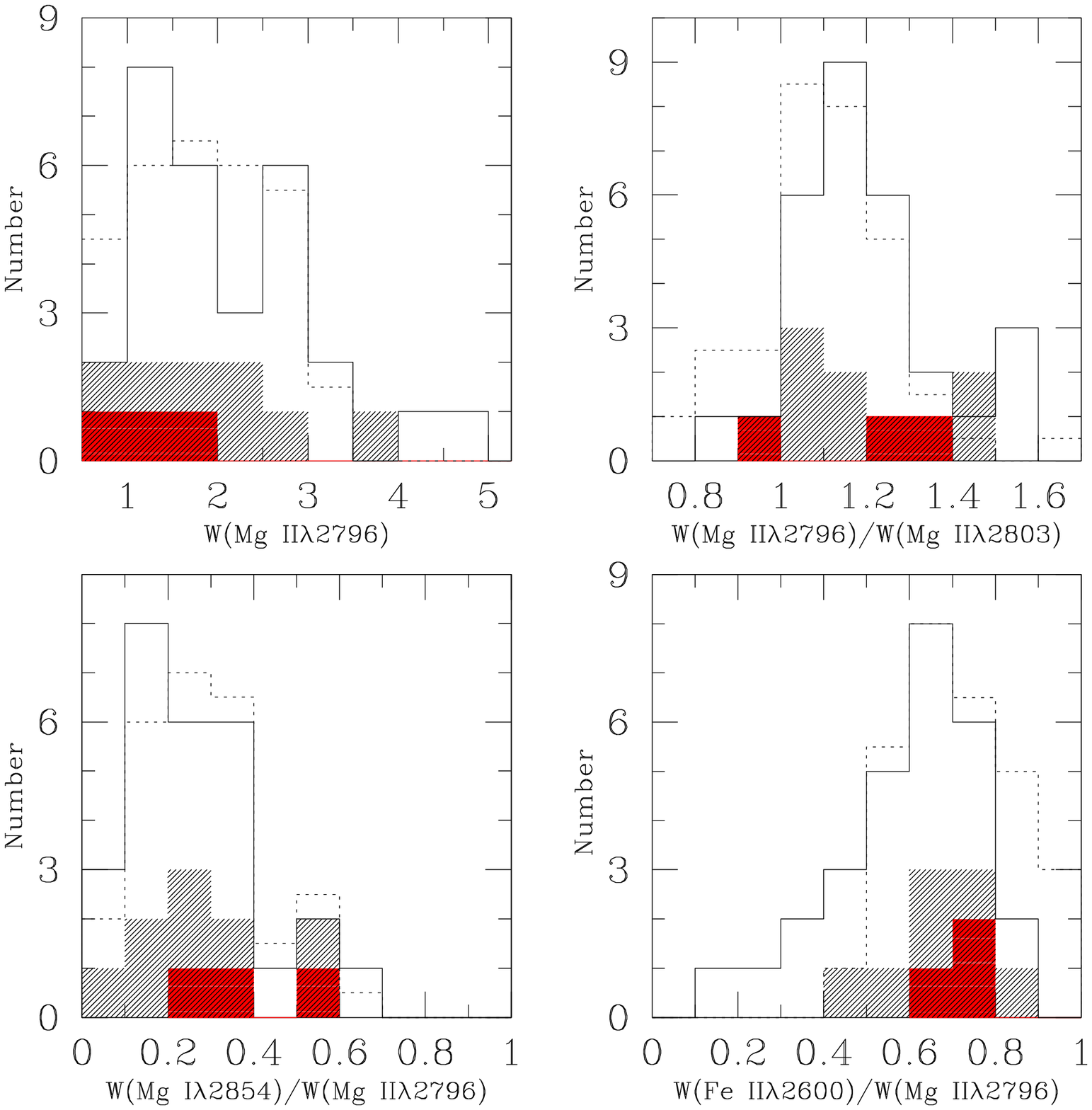}}
\caption{Distributions of W(Mg~{\sc ii}) and different line ratios.
Solid and dotted histograms are respectively for our GMRT sample and the
sample of DLAs (downscaled by factor 2) studied by Rao \& Turnshek (2000).  Light and dark
 histograms are for the systems observed for 21-cm absorption so far and the systems with detections respectively.}
\label{Fig4}
\end{figure}

We also report upper limits on 7 systems with one of them being a tentative
detection. We find a detection rate of 21-cm absorption of roughly 
30\% in our survey so far. This is much higher than that achieved earlier 
(see for example Lane, 2000), mainly because we have restricted 
ourselves to the stronger Mg~{\sc ii} systems. Since the objects 
observed during the first-phase of the survey are drawn randomly from 
the parent sample, most probably the detection rate of our whole survey 
will be close to what we have achieved so far.
In Fig.~\ref{Fig4} we compare the properties of our systems with those of
DLAs in the sample of Rao \& Turnshek (2000). The distribution of different 
equivalent width ratios together with the distribution
of W(Mg~{\sc ii}) in our sample is consistent with what is observed in 
the DLA population. 

It is important to obtain independent constraints on the time variation of 
fundamental constants in order to resolve the controversy regarding the 
variation of $\alpha$ (Murphy et al. 2003; Srianand et al. 2004).  As the 
energy of the 21-cm transition depends on the electron-to-proton
mass ratio ($\mu$), the fine-structure constant ($\alpha$) and the proton
G-factor (G$_p$), high resolution optical spectra in conjunction with
high resolution 21-cm spectra can be used to probe the variation of
these constants (Tzanavaris et al. 2005). However, it is important first to 
understand the origin of the relative shifts that we observe between the 
redshifts of the atomic heavy element lines and the 21-cm absorption. This
needs detailed modeling of the absorption systems taking into account all
transitions simultaneously. This is what we plan to do in the near 
future.

%%%%%%%%%%%%%%%%%%%%%%%%%%%%%%%%%%%%%%%%%%%%%%%%%%%%%
\acknowledgements{
RS and PPJ gratefully acknowledge support from the Indo-French Centre
for the Promotion of Advanced Research (IFCPAR) under contract No. 3004-3.  
We thank the referee for useful comments, Rajaram Nityananda for
encouragement and GMRT staff for their co-operation during observations.
The GMRT facility is run by NCRA-TIFR.
}

\clearpage

\begin{table}
\caption{Observational log and results.}
\begin{center}
\begin{tabular} {lclccr}
\hline
\hline
\multicolumn{1}{c}{Name}& \zabs      &Time               &Peak &rms$^b$& $\tau^c$\\
                        &            & (hr)              &flux$^a$&    &         \\
\hline                                                    
J0108$-$0037            &  1.3710    &  4.4              &1276 & 2.9   &    0.070\\
J0214$+$1405            &  1.4463    &  4.4              &220  & 2.7   & $<$0.012\\
J0240$-$2309            & 1.3647     &  8.0              &5100&   5.2  & $<$0.001\\
J0748$+$3006            &  1.4470    &  5.5              &347  & 3.5   & $<$0.010\\
J0804$+$3012            &  1.1908    &  6.0              &2050 & 1.9   &    0.006\\
                        &            &  6.2$^d$          &2087 & 1.6   &    0.006\\
J0845$+$4257            &  1.1147    &  5.0              &224  & 3.0   & $<$0.013\\
J1017$+$5356            &  1.3055    &  4.4              &127  & 2.3   & $<$0.018\\
J1411$-$0300$^e$        &  1.4160    &  3.9              &244  & 3.0   & $<$0.012\\
                        &            &                   &161  & 3.0   & $<$0.019\\
J1604$-$0019            &  1.3245    &  3.3$^{d,f}$      &375  & 3.9   & $<$0.010\\
J2358$-$1020            &  1.1726    &  3.9$^{d,f}$      &443  & 1.4   &    0.033\\
                        &            &  5.5$^g$          &420  & 2.5   &    0.035\\
\hline
\multicolumn{6}{l}{\small $^a$ peak flux (mJy/beam); $^b$ spectral rms (mJy/beam/channel)}\\
\multicolumn{6}{l}{\small $^c$ peak optical depth or 1-$\sigma$ limit per channel;}\\
\multicolumn{6}{l}{\small $^d$ 2 MHz band width.}\\
\multicolumn{6}{l}{\small $^e$ The two spectra correspond to peaks P$_1$ and P$_2$.}\\
\multicolumn{6}{l}{\small $^f$ 256 channels used; $^g$ 0.5 MHz bandwidth}\\
\end{tabular}
\end{center}
\label{obslog}
\end{table}

\clearpage

\clearpage
  
\clearpage

\clearpage

\clearpage

\begin{thebibliography}{99}
%
\bibitem[]{484} Beasley, A. J., Gordon, D., Peck, A. B., Petrov, L., MacMillan, D. S., 
Fomalont, E. B., \&  Ma, C., 2002, ApJS, 141, 13
%
\bibitem{brg83} Briggs, F. H., \& Wolfe, A.M., 1983, ApJ, 268, 76
%
\bibitem{cur05} Curran, S.J., Murphy, M. T.,  Pihlstr\"om, V. M., Webb, J. K, \& 
Purcell, C. R. 2005, MNRAS, 356, 1509
%
\bibitem{cur06} Curran, S. J., \& Webb, J. 2006, MNRAS, 371, 356.
%
\bibitem[]{493} Fomalont, E. B., Frey, S., Paragi, Z., Gurvits, L. I., Scott, W. K., 
Taylor, A. R., Edward, P. G., \& Hirabayashi, H, 2000, ApJS, 131, 95
%
\bibitem{gup06} Gupta, N., Salter, C.J., Saikia, D.J., Ghosh, T., \& Jeyakumar, S., 
2006, MNRAS, in press (astro-ph/0605423)
%
\bibitem{hen06} Heinm\"uller, J., Petitjean, P., Ledoux, C., Caucci, S., \&
Srianand, R., 2006, A\&A, 449, 33
%
\bibitem{kan03} Kanekar, N., \& Chengalur, J.N., 2003, A\&A, 399, 857
%
\bibitem{kan06} Kanekar, N., Subrahmanyan, R., Ellison, S.L., Lane, W.M., 
\& Chengalur, J.N., 2006, MNRAS, 370, L46
%
\bibitem{lan00} Lane W., 2000, PhD Thesis, University of Groningen
%
\bibitem{lan87} Lanzetta, K. M., Wolfe, A. M., \& Turnshek, D. A, 1987, ApJ, 322, 739 
%
\bibitem{led03} Ledoux, C., Petitjean, P., \& Srianand, R., 2003, MNRAS, 346, 209
%
\bibitem{mol04} M\"oller, P., Fynbo, J. P. U., \& Fall, S. M., 2004, A\&A, 422, 33
%
\bibitem{mur03} Murphy, M. T., Webb, J. K.,\& Flambaum, V. V., 2003, MNRAS, 345, 609
%
\bibitem{prc06} Prochter, G. E., Prochaska, J. X., Burles, S. M., 2006, ApJ, 639, 766
%
\bibitem{rao00} Rao, S.M., Turnshek, D.A., 2000, ApJS, 130, 1
%
\bibitem[]{517} Rodr\' iguez, E., Petitjean, P., Aracil, B., Ledoux, C., \&
Srianand, R. 2006, A\&A, 446, 791
%
\bibitem[]{519} Srianand, R., Chand, H., Petitjean, P., \& Aracil, B, 2004, PRL, 92, 121302
%
\bibitem[]{521} Srianand, R., Gupta, N., \& Petitjean, P., 2006, MNRAS, in press (astro-ph/0611327)
%
\bibitem{sri05} Srianand, R., Petitjean, P., Ledoux, C., Ferland, G., \& Shaw, G., 2005, 
MNRAS, 362, 549
%
\bibitem[]{525} Tzanavaris, P., Webb, J. K., Murphy, M. T., Flambaum, V. V., \& Curran, S. J., 
2005, PRL, 95, 1301
%
\bibitem{wlf04} Wolfe, A.M., Howk, J.C., Gawiser, E., Prochaska, J.X., \& Lopez, S., 
2004, ApJ, 615, 625
%
\bibitem{wlf05} Wolfe, A.M., Gawiser, E., Prochaska, J.X., 2005, ARA\&A, 43, 861 
%
\bibitem{York06} York, D. G., et al. 2006, MNRAS, 367, 945.
\end{thebibliography}
\end{document}